\documentclass[aip,jap,reprint, floatfix]{revtex4-1}
\pdfoutput=1
\usepackage{graphicx}
\usepackage{bm}
\usepackage{siunitx}
\usepackage{amsmath}
\usepackage[colorlinks=true, allcolors=blue]{hyperref}

\newcommand{\vect}[1]{\bm{\mathrm{#1}}}

\newcommand{\subl}[2][]{^{\mathrm{(#2)}#1}}

\begin{document}

\title{Micromagnetic simulation of exchange coupled ferri-/ferromagnetic composite in bit patterned media}
\author{Harald Oezelt}
\email{harald.oezelt@fhstp.ac.at}
\thanks{The following article appeared in H. Oezelt et al., Journal of Applied Physics \textbf{117}, 17E501 and may be found at \url{http://dx.doi.org/10.1063/1.4906288}. This article may be downloaded for personal use only. Any other use requires prior permission of the author and AIP Publishing. Copyright (2015) American Institute of Physics.}
\affiliation{Industrial Simulation, University of Applied Sciences, 3100 St. P\"olten, Austria}
\author{Alexander Kovacs}
\affiliation{Industrial Simulation, University of Applied Sciences, 3100 St. P\"olten, Austria}
\author{Phillip Wohlh\"uter}
\affiliation{Laboratory for Mesoscopic Systems, Department of Materials, ETH Zurich, 8093 Zurich, Switzerland}
\affiliation{Laboratory for Micro- and Nanotechnology, Paul Scherrer Institute, 5232 Villigen PSI, Switzerland}
\author{Eugenie Kirk}
\affiliation{Laboratory for Mesoscopic Systems, Department of Materials, ETH Zurich, 8093 Zurich, Switzerland}
\affiliation{Laboratory for Micro- and Nanotechnology, Paul Scherrer Institute, 5232 Villigen PSI, Switzerland}
\author{Dennis Nissen}
\affiliation{Institute of Physics, Chemnitz University of Technology, 09126 Chemnitz, Germany}
\affiliation{Institute of Physics, University of Augsburg, 86159 Augsburg, Germany}
\author{Patrick Matthes}
\affiliation{Institute of Physics, University of Augsburg, 86159 Augsburg, Germany}
\author{Laura Jane Heyderman}
\affiliation{Laboratory for Mesoscopic Systems, Department of Materials, ETH Zurich, 8093 Zurich, Switzerland}
\affiliation{Laboratory for Micro- and Nanotechnology, Paul Scherrer Institute, 5232 Villigen PSI, Switzerland}
\author{Manfred Albrecht}
\affiliation{Institute of Physics, Chemnitz University of Technology, 09126 Chemnitz, Germany}
\affiliation{Institute of Physics, University of Augsburg, 86159 Augsburg, Germany}
\author{Thomas Schrefl}
\affiliation{Industrial Simulation, University of Applied Sciences, 3100 St. P\"olten, Austria}
\affiliation{Centre for Integrated Sensor Systems, Danube University Krems, 2700 Wiener Neustadt, Austria}




\begin{abstract}
Ferri-/ferromagnetic exchange coupled composites are promising candidates for bit patterned media because of the ability to control the magnetic properties of the ferrimagnet by its composition. A micromagnetic model for the bilayer system is presented where we also incorporate the microstructural features of both layers. Micromagnetic finite element simulations are performed to investigate the magnetization reversal behaviour of such media. By adding the exchange coupled ferrimagnet to the ferromagnet, the switching field could be reduced by up to $\SI{40}{\%}$ and also the switching field distribution is narrowed. To reach these significant improvements, an interface exchange coupling strength of $\SI{2}{mJ/m^2}$ is required.
\end{abstract}


\maketitle

\section{Introduction}
In the attempt to move forward to higher data density in magnetic storage devices, the concept of bit patterned media is the next logical step. In order to maintain thermal stability of the bits and keep them writeable at the same time, Suess et al.~\cite{Suess2007} and Victora et al.~\cite{Victora2005} proposed the idea of exchange spring media. This approach allows the use of material with high magnetic anisotropy by decreasing the required switching field with an exchange-coupled soft magnetic material. Experimental studies confirmed the feasibility of exchange coupled media in multilayer structures~\cite{Casoli2008} and later also in bit patterned media~\cite{Goh2009, Sbiaa2009, Hauet2009}. Krone et al.~\cite{Krone2010} performed micromagnetic simulations of arrays consisting of exchange coupled composite stacks and also graded media, where the magnetic anisotropy constant was decreased quadratically across ten layers. 

In this paper we investigate exchange coupled bilayer dots where a ferrimagnetic material, such as FeTb or FeGd, represents the soft magnetic layer. Ferrimagnetic materials have been extensively studied~\cite{Giles1991, Mansuripur1988a} and used as magneto-optical recording media~\cite{Kryder1985, Jenkins2003}. A big advantage of using ferrimagnetic layers is the possibility to tailor their magnetic properties through their composition with respect to the desired working temperature~\cite{Mimura1978}. Moreover, since these layers are amorphous, the lack of crystalline defects may positively influence the switching field distribution of the exchange coupled ferromagnetic layer. 

In the following, we will describe the micromagnetic model used for the exchange coupled ferri-/ferromagnetic composite dots. We also consider the granular structure of the magnetically harder ferromagnet and material inhomogeneities in the softer, amorphous ferrimagnet in our geometrical model. Then, micromagnetic finite element simulations are used to calculate the magnetization reversal of dots with varying microstructure, diameter and interface coupling strength. This in turn enables the investigation of the magnetization configuration during reversal and the switching field distribution.

\section{Micromagnetic Model}
In this study we consider cylindrical dots composed of a ferromagnetic layer $\Omega^{\mathrm{FM}}$ and a ferrimagnetic layer $\Omega^{\mathrm{FI}}$ collinearly exchange-coupled at the interface $\Gamma$. Since a detailed explanation of a ferri-/ferromagnetic bilayer model has already been given in our previous work~\cite{Oezelt2015}, we provide a brief summary here. 

\begin{figure}[htb]
\centering
\includegraphics[width=\columnwidth]{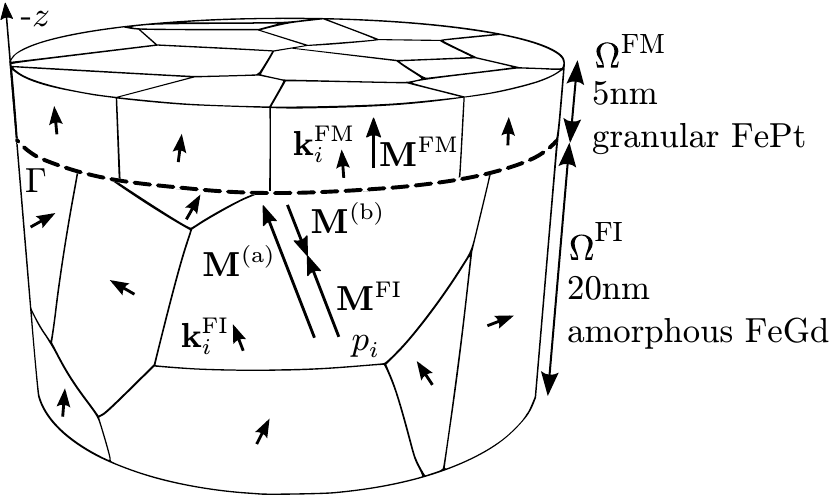}
\caption{Geometric model of the bilayer dot with a ferrimagnetic phase $\Omega^{\mathrm{FI}}$ and a ferromagnetic phase $\Omega^{\mathrm{FM}}$ connected at the interface~$\Gamma$. The magneticaly softer, amorphous $\Omega^{\mathrm{FI}}$ is divided in patches $p_i$ with varying uniaxial anisotropic properties $\vect{k}_i^{\mathrm{FI}}$ and $K_{u,i}^{\mathrm{FI}}$. The granular hard magnetic phase, $\Omega^{\mathrm{FM}}$, posseses strong out-of-plane uniaxial anisotropy.\label{fig:modelexp}}
\end{figure}

While the finite element simulation of ferromagnets is a common task, the mathematical model for ferrimagnets has to be adapted. We follow Mansuripur's~\cite{Mansuripur95} approach and assume that the ferrimagnetic sublattices are strongly coupled antiparallel at all times. Therefore we can substitute the magnetic moments $\vect{M}\subl{a}$, $\vect{M}\subl{b}$ of the sublattices with an effective net moment $\vect{M}^{\mathrm{FI}}$ by defining its net magnitude as $M^{\mathrm{FI}}=M\subl{a}-M\subl{b}$ and its unit vector as $\vect{m}=\vect{m}\subl{a}=-\vect{m}\subl{b}$ (see FIG.~\ref{fig:modelexp}). The Gilbert equations of both sublattices can then be summed up to obtain an effective Gilbert equation. Since we are only interested in the static hysteresis behaviour, the damping constant is set to $\alpha_{\mathrm{eff}}=1$. 

To take into account the exchange coupling between the two layers, we extend the equation for the effective field of each layer with the interface exchange field $\vect{H}_{\mathrm{ixhg}}=-1/\mu_0\,\delta E_{\mathrm{ixhg}}/\delta \vect{M}$ exerted by the respective neighbouring layer. The exchange energy across the interface $\Gamma$ is given by $E_{\mathrm{ixhg}}=-A_{\Gamma} / a \int_{\Gamma}\vect{u}^{\mathrm{FM}}\vect{u}^{\mathrm{FI}}\,d\Gamma$, where $A_{\Gamma}$ is the exchange stiffness constant, $a$ is the distance between spins in a simple cubic lattice and $\vect{u}$ is the unit vector of each spin direction. Due to the microstructural differences both layers have to be meshed separately. As the mesh nodes at the interface do not match, we employ a surface integral technique and calculate $E_{\mathrm{ixhg}}$ by using a symmetric Gaussian quadrature rule for triangles~\cite{Dunavant1985, Dean2010}.

The geometrical model used for the simulations is depicted in FIG.~\ref{fig:modelexp}. The $\Omega^{\mathrm{FM}}$ model is a $\SI{5}{nm}$ thick, $L1_0$ chemically ordered $\mathrm{Fe}_{52}\mathrm{Pt}_{48}$ layer with an average grain diameter of $\SI{13}{nm}$. The layer exhibits a saturation polarization of $J_{\mathrm{s}}^{\mathrm{FM}}=\SI{1.257}{T}$ and an exchange stiffness constant of $A_{\mathrm{x}}^{\mathrm{FM}}=\SI{10}{pJ/m}$. Each grain has its own randomized anisotropic constant and uniaxial anisotropic direction. The average assigned anisotropic constant is $K_{\mathrm{u}}^{\mathrm{FM}}=\SI{1.3}{MJ/m^3}$ with a standard deviation of $0.05K_{\mathrm{u}}^{\mathrm{FM}}\,\si{J/m^3}$. The uniaxial anisotropic direction is limited within a cone angle of $\ang{15}$ from the out-of-plane ($z$-) axis. No intergrain phase is considered.

The $\Omega^{\mathrm{FI}}$ phase is an amorphous, $\SI{20}{nm}$ thick $\mathrm{Fe}_{74}\mathrm{Gd}_{26}$ layer. The ferrimagnet is characterized by a saturation polarization of $J_{\mathrm{s}}^{\mathrm{FI}}=\SI{0.268}{T}$, an exchange stiffness constant of $A_{\mathrm{x}}^{\mathrm{FI}}=\SI{2}{pJ/m}$. To incorporate material inhomogeneities in the amorphous model we divide the layer into patches $p_i$ with an average diameter of $\SI{13}{nm}$ as suggested by Mansuripur and Giles~\cite{Mansuripur1991}. Each patch exhibits its own randomly assigned anisotropic constant and uniaxial anisotropic direction. The average anisotropic constant is $K_{\mathrm{u}}^{\mathrm{FI}}=\SI{10}{kJ/m^3}$ with a standard deviation of $0.2K_{\mathrm{u}}^{\mathrm{FI}}\,\si{J/m^3}$. The uniaxial anisotropic direction varies within a cone angle of $\ang{90}$ from patch to patch.

The micromagnetic simulations are performed by using the finite element micromagnetic package FEMME~\cite{Schrefl2007}. We investigate the dependence of reversal curve and especially the switching field $H_{\mathrm{sw}}$ on the dot diameter and the exchange coupling at the interface. In order to calculate the switching field distribution, 20 simulation runs for each dot diameter and interface coupling strength were performed. Each simulation had its individual mesh for both layers with randomized microstructure generated by the software Neper~\cite{Quey2011}. Also the randomized anisotropic properties of both layers were generated anew for each simulation within the limits described previously. The mesh size for both layers was set to $\SI{2}{nm}$.

\section{Results and Discussion}
\begin{figure}[htb]
\centering
\includegraphics[width=\columnwidth]{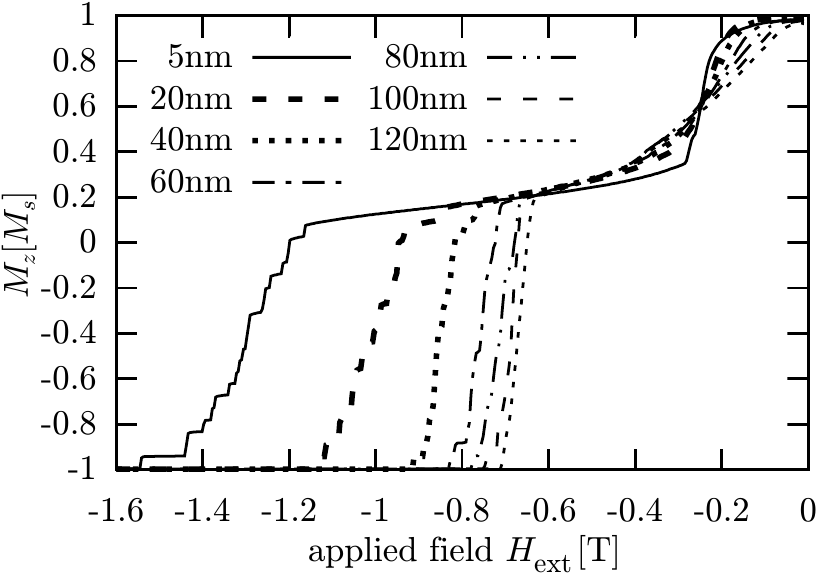}
\caption{Reversal curves of dot arrays for different dot diameter. The layers of the dot are strongly coupled with $A_{\Gamma} / a=\SI{5}{mJ/m^2}$.\label{fig:hysteresis}}
\end{figure}

By applying an increasing external field $H_{\mathrm{ext}}$ to the fully saturated dot in the opposite direction ($-z$) to the magnetization, the reversal curve is computed. For each set of dot diameter and interface exchange strength we compute the average reversal curve over the 20 randomized simulations. So the result can be seen as the reversal curve of an array of 20 dots of equal diameter, but varying microstructure and anisotropic properties. The averaged reversal curves for different dot diameters with an exchange coupling strength of $A_{\Gamma} / a=\SI{5}{mJ/m^2}$ at the interface are depicted in FIG.~\ref{fig:hysteresis}. 

\begin{figure}[htb]
\centering
\includegraphics[width=\columnwidth]{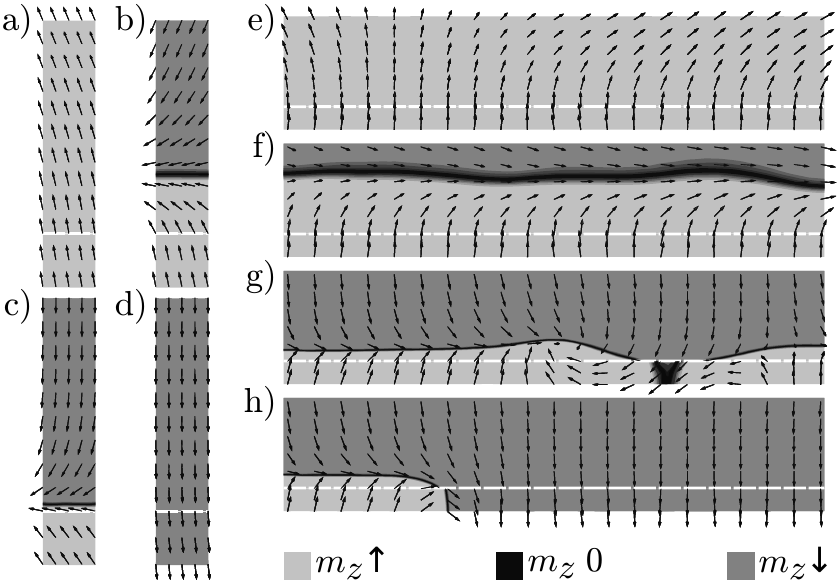}
\caption{Reversal process of a dot with $d=\SI{5}{nm}$ from a) to d) and of a dot with  $d=\SI{120}{nm}$ from e) to h). The interface $\Gamma$ is the white dashed line while the $\Omega^{\mathrm{FI}}$ is the upper and the $\Omega^{\mathrm{FM}}$ is the lower layer.\label{fig:reversal}}
\end{figure}

For all diameters the soft magnetic $\Omega^{\mathrm{FI}}$ phase switches at about $\SI{-0.27}{T}$. With increasing diameter the reversal curve of the soft magnetic phase gets flattened. This can be accredited to the shape anisotropy, since the layer thickness is fixed for all models. This can also be seen in FIG.~\ref{fig:reversal} where, in contrast to the $\SI{5}{nm}$ dot, the $\SI{120}{nm}$ dot shows an inhomogeneous reversal of the magnetic moments starting with an in-plane configuration at the surface of the ferrimagnet (FIG.~\ref{fig:reversal}f). The reversal of the hard magnetic $\Omega^{\mathrm{FM}}$ phase in FIG.~\ref{fig:reversal} strongly depends on the dot diameter. With smaller diameters the $\Omega^{\mathrm{FM}}$ phase consists only of one or a few grains, which leads to a flattened reversal curve when averaged over the 20 simulations, i. e. a broader switching field distribution. The switching field $H_{\mathrm{sw}}^{\mathrm{FM}}$ drastically increases with decreasing diameter. This is because with increasing diameter the model changes from the single domain to a multi domain regime. 

In FIG.~\ref{fig:reversal} the magnetization configuration during reversal of the bilayer is depicted on an $x$-$z$-slice through the center of a $d=\SI{5}{nm}$ and a $d=\SI{120}{nm}$ dot. The regions in bright gray are still not reversed, the dark gray areas are already reversed, whereas the domain walls are in black. 

By applying the external field in down direction in the small dot, the magnetic moments of the upper region coherently switch and form a domain wall due to the exchange coupling at the interface (FIG.~\ref{fig:reversal}b). With increasing field the domain wall gets pushed towards the interface c), when eventually the $\Omega^{\mathrm{FM}}$ phase nucleates as a single domain d). In the $\SI{120}{nm}$ dot the $\Omega^{\mathrm{FI}}$ phase starts to rotate more inhomogeneously e) and turns in-plane at the surface f). At this state the domain wall is widened because of its $\ang{90}$ configuration. With increasing $H_{\mathrm{ext}}$ the domain wall gets narrower and is pushed through the interface into the $\Omega^{\mathrm{FM}}$ phase g). Compared to the smaller dot, the reversal of the harder phase is much more inhomogeneous and a lateral domain wall movement can be observed leading to full reversal h). 

\begin{figure}[htb]
\centering
\includegraphics[width=\columnwidth]{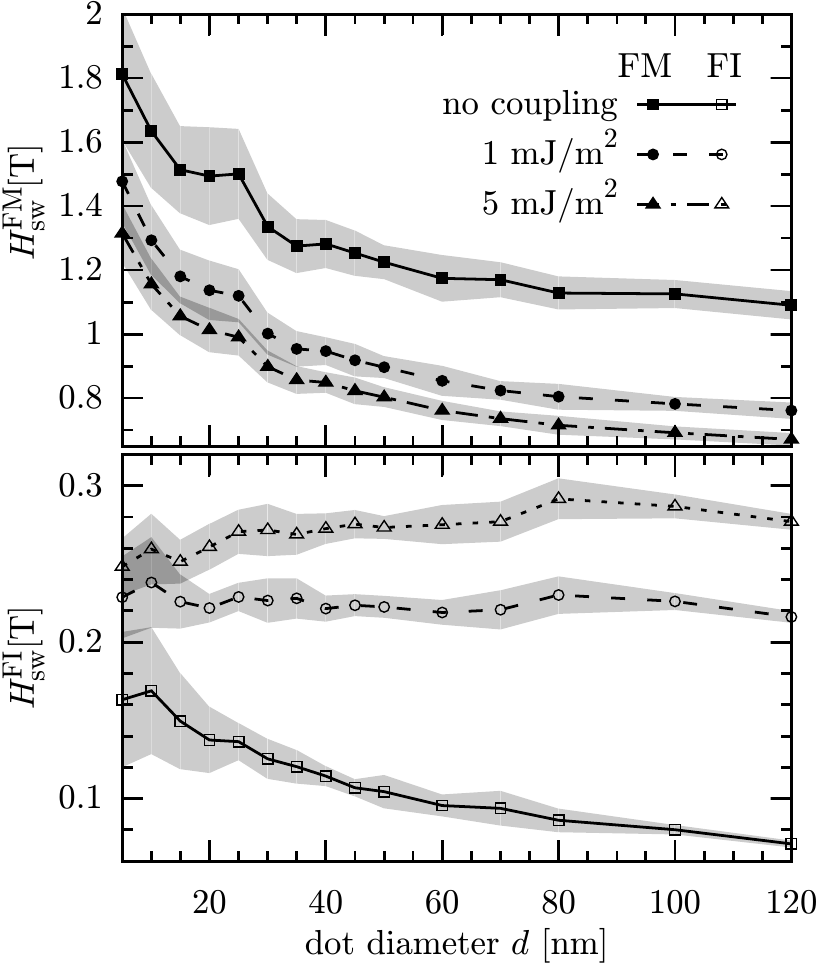}
\caption{Averaged switching field of the $\Omega^{\mathrm{FM}}$ and the  $\Omega^{\mathrm{FI}}$ phase for different interface coupling strengths depending on the dot diameter. The standard deviation $\sigma_{\mathrm{sw}}$ of 20 simulations for each data point is depicted as the gray area $H_{\mathrm{sw}}\pm\sigma_{\mathrm{sw}}$.\label{fig:diahsStdev}}
\end{figure}

This translation from homogeneous to inhomogeneous reversal of the hard magnetic phase can be clearly recognized in FIG.~\ref{fig:diahsStdev} where the $H^{\mathrm{FM}}_{\mathrm{sw}}$ curves drop between $\SI{25}{nm}$ and $\SI{30}{nm}$ dot diameter. In FIG.~\ref{fig:diahsStdev} we show the switching fields of both layers, again averaged over the 20 randomized simulation runs. The switching fields, $H^{\mathrm{FM}}_{\mathrm{sw}}$ in the upper and $H^{\mathrm{FI}}_{\mathrm{sw}}$ in the lower area, are defined as $M_z^{\mathrm{FI}}(H^{\mathrm{FI}}_{\mathrm{sw}})=0$ and $M_z^{\mathrm{FM}}(H^{\mathrm{FM}}_{\mathrm{sw}})=0$. The curves for three different interface coupling strengths are shown. Additionally the standard deviation of the 20 simulation runs for each data point is shown as a grey shade: $H_{\mathrm{sw}}\pm\sigma_{\mathrm{sw}}$.

Without a coupled ferrimagnet, $H^{\mathrm{FM}}_{\mathrm{sw}}$ is reduced by $\SI{39}{\%}$, when moving from a $\SI{5}{nm}$ to a $\SI{120}{nm}$ dot diameter. For a strongly coupled bilayer this reduction is improved to $\SI{50}{\%}$. If  we look at a specific dot diameter, introducing the coupled ferrimagnet reduces $H^{\mathrm{FM}}_{\mathrm{sw}}$ by $\SI{30}{\%}$ to $\SI{40}{\%}$. The higher the diameter, the higher the reduction of the switching field of the hard phase.
While the switching field of a single ferrimagnetic layer would decrease with growing diameter, an increasing interface coupling can stabilize or even cause an increase of $H^{\mathrm{FI}}_{\mathrm{sw}}$ by about $\SI{12}{\%}$ within the investigated diameter range.
FIG.~\ref{fig:diahsStdev} also shows that the switching field distribution decreases with increasing diameter and interface exchange coupling for the ferromagnet. The ferrimagnetic phase shows a significant reduction of the switching field distribution when increasing the diameter from $5$ to $\SI{20}{nm}$. 

\begin{figure}[htb]
\centering
\includegraphics[width=\columnwidth]{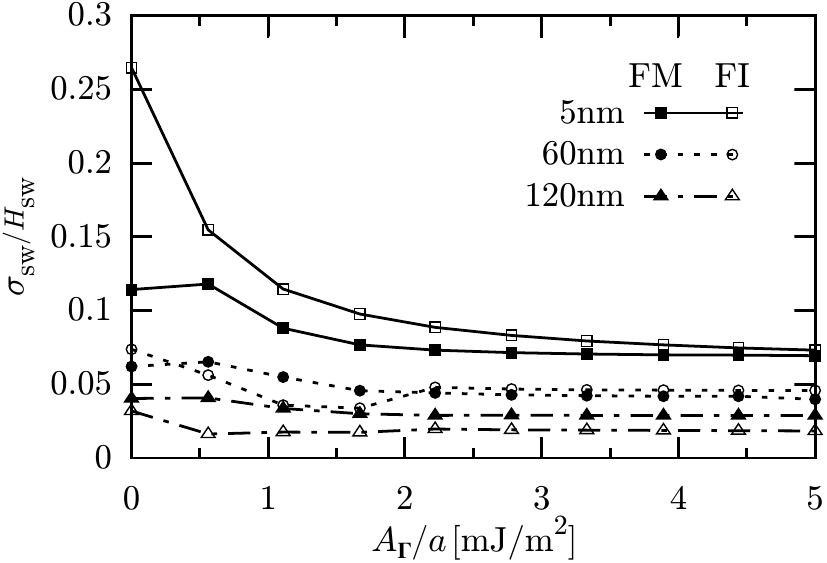}
\caption{Relative standard deviation of the switching field of both layers for different dot diameters as a function of interface exchange coupling strength.\label{fig:Stdevixhg}}
\end{figure}

This behaviour can also be seen in FIG.~\ref{fig:Stdevixhg} where the relative standard deviation of the switching field $\sigma_{\mathrm{sw}}/H_{\mathrm{sw}}$ is plotted against the exchange coupling strength. The solid symbols refer to the ferromagnetic phase and the empty symbols to the ferrimagnetic phase for three different dot diameters.

Coupling the ferrimagnetic layer to the ferromagnet decreases the relative standard deviation $\sigma^{\mathrm{FM}}_{\mathrm{sw}}/H^{\mathrm{FM}}_{\mathrm{sw}}$ from  $\SI{11}{\%}$ to below $\SI{7}{\%}$ for $\SI{5}{nm}$ diameter. For larger diameters it decreases to $\SI{4}{\%}$ for $\SI{60}{nm}$ or even below $\SI{2}{\%}$ for $\SI{120}{nm}$. The relative standard deviation for the ferrimagnet is also reduced with increasing interface exchange energy, especially for the $\SI{5}{nm}$ diameter dot, where it is reduced from $\SI{26}{\%}$ to $\SI{7}{\%}$. The major change of the switching field distribution occurs below $A_{\Gamma} / a=\SI{2}{mJ/m^2}$ and only slightly improves above.

\section{Summary}
A micromagnetic model for exchange coupled ferri-/ferromagnetic bilayer dots was presented. We performed a series of simulations for the dots of diameters from $\SI{5}{nm}$ to $\SI{120}{nm}$ with varying interface exchange coupling strength from $0$ to $\SI{5}{mJ/m^2}$. For each parameter set, 20 simulations were performed with randomized microstructure and anisotropic properties.

We found that with increasing dot diameter the switching field of the hard phase drastically decreases and also narrows the switching field distribution. Dots with small diameters exhibit homogeneous switching behaviour, only interrupted when in the ferrimagnet a domain wall close to the exchange coupled interface is created and is slowly pushed towards it. Dots with larger diameters reverse more inhomogeneously, building an in-plane orientation configuration and show a lateral domain wall movement in the hard magnetic phase. The switching field and its distribution can also be controlled by the exchange coupling strength at the interface. With increasing exchange coupling, the ferromagnetic switching field is reduced by $\SI{30}{\%}$ for $\SI{5}{nm}$ dots and $\SI{40}{\%}$ for $\SI{120}{nm}$ dots. Its distribution is improved to $\sigma^{\mathrm{FM}}_{\mathrm{sw}}/H^{\mathrm{FM}}_{\mathrm{sw}}=\SI{7}{\%}$ for $\SI{5}{nm}$ dots and $\sigma^{\mathrm{FM}}_{\mathrm{sw}}/H^{\mathrm{FM}}_{\mathrm{sw}}=\SI{2}{\%}$ for $\SI{120}{nm}$ dots. To reach significant improvements, an interface exchange coupling strength of $A_{\Gamma} / a=\SI{2}{mJ/m^2}$ is required.

\begin{acknowledgments}
We gratefully acknowledge the financial support provided by the Austrian Science Fund (FWF Grant No. I821), the German Research Foundation (DFG Grant No. AL 618/17-1) and the Swiss National Science Foundation (SNF Grant No. 200021L\_137509).
\end{acknowledgments}

\bibliography{references}
\end{document}